# Superconductivity in $Mo_4Ga_{20}As$ with Endohedral Gallium Clusters

Bin-Bin Ruan[1][*], Le-Wei Chen[1,2], Yun-Qing Shi[1,2], Jun-Kun Yi[1,2], Qing-Song Yang[1,2], Meng-Hu Zhou[1], Ming-Wei Ma[1], Gen-Fu Chen[1,2] and Zhi-An Ren[1,2][*]

[1] Institute of Physics and Beijing National Laboratory for Condensed Matter Physics, Chinese Academy of Sciences, Beijing 100190, China
[2] School of Physical Sciences, University of Chinese Academy of Sciences, Beijing 100049, China

[*] Corresponding authors (email: bbruan@mail.ustc.edu.cn; renzhian@iphy.ac.cn)

**Abstract**
We report the discovery and detailed investigation of superconductivity in $Mo_4Ga_{20}As$. $Mo_4Ga_{20}As$ crystallizes in the space group of $I4/m$ (No. 87), with lattice parameters $a$ = 12.86352 Å and $c$ = 5.30031 Å. The resistivity, magnetization, and specific heat data reveal $Mo_4Ga_{20}As$ to be a type-II superconductor with $T_c$ = 5.6 K. The upper and lower critical fields are estimated to be 2.78 T and 22.0 mT, respectively. In addition, electron-phonon coupling in $Mo_4Ga_{20}As$ is possibly stronger than the BCS weak-coupling limit. First-principles calculations suggest the Fermi level being dominated by the Mo-4$d$ and Ga-4$p$ orbitals.

## 1. Introduction

Searching for new superconductors with high transition temperature ($T_c$) has long been a key issue in condensed matter physics. Despite the lack of a universal theory for superconductivity, researchers have developed empirical guidelines for material design. Certain crystal structures (*e.g.* $A$15-type, structures with $CuO_2$/$Fe_2As_2$ layers) are favorable [1-3]. Besides, incorporation of some elements, such as niobium, boron, or hydrogen was found to be beneficial to the occurrence of superconductivity, as shown in $Nb_3Sn$ [4], $MgB_2$ [5], and hydrides [6].

In recent years, endohedral cluster superconductors have attracted much attention due to their chemical diversity as well as intriguing physical properties [7]. For instance, unconventional superconductivity was revealed in $PuCoGa_5$ ($T_c$ = 18.0 K) [8] and $CeCoIn_5$ ($T_c$ = 2.3 K) [9]. As for the binary compounds, $Rh_2Ga_9$ ($T_c$ = 1.9 K) and $Ir_2Ga_9$ ($T_c$ = 2.2 K) were found to host noncentrosymmetric crystal structures [10], $RuAl_6$ ($T_c$ = 1.2 K) was proposed to be a topological crystalline insulator [11], and strong electron-phonon coupling was observed in Mo-based compounds such as $Mo_8Ga_{41}$ ($T_c$ = 9.8 K) [12], $Mo_6Ga_{31}$ ($T_c$ = 8.2 K) [13], $Mo_4Ga_{18.7}Sn_{1.824}$ ($T_c$ = 5.8 K) [14], and $Mo_4Ga_{20}Sb$ ($T_c$ = 6.6 K) [15]. Notice that $Mo_8Ga_{41}$ hosts a fairly high $T_c$, which is also the highest among the binary endohedral cluster compounds. Moreover, muon spin rotation/relaxation measurements revealed evidences of multigap in $Mo_8Ga_{41}$ [16].

Due to the apparently higher $T_c$ and rich physics, the Mo-based endohedral cluster superconductors are of special interests. Among them, the $Mo_4Ga_{21}$-type superconductors, the topic of this work, form a large structural family. Although the binary "$Mo_4Ga_{21}$" was never obtained, the structure could be stabilized with a third element $E$, forming $Mo_4Ga_{20}E$ phases ($E$ = Sn, Sb, S, Se, Te) [14,15,17]. $Mo_4Ga_{20}E$ were basically stoichiometric compounds, with $E$ taking the Ga sites in an ordered manner, while partial occupancies were observed in $Mo_4Ga_{20.38}S_{0.62}$ and $Mo_4Ga_{18.7}Sn_{1.824}$ [14,15]. All of the compounds reported in $Mo_4Ga_{20}E$ family are superconductors, with $T_c$ ranging from 3.3 to 6.6 K [14,15].

Here we report the discovery of superconductivity in $Mo_4Ga_{20}As$ ($T_c$ = 5.6 K), another member of the $Mo_4Ga_{20}E$ family. The superconducting properties were investigated in detail with electrical resistivity, magnetization, and specific heat measurements. The superconducting parameters were determined. First-principles calculations revealed similarities between $Mo_4Ga_{20}As$ and previously reported isostructural compounds.

## 2. Methods

Polycrystalline samples of $Mo_4Ga_{20}As$ were synthesized by solid state reactions with starting materials: Mo (99.95%, powder), Ga (99.99%, ingot), and As (99.999%, powder). Stoichiometric amounts of the starting materials were mixed, placed into alumina crucibles, and heated at 973 K for 48 hours in vacuum-sealed quartz tubes. The products were then ground into fine powders, pressed into pellets, sealed into quartz tubes, and annealed at 973 K for 5 days. The sample manipulation was conducted in a glove box filled with high-purity argon. The final products had greyish metallic lusters and were stable in air. We also synthesized samples with nominal compositions of $Mo_4Ga_{21-x}As_x$ with $x \neq 1.0$. However, when $x$ differed from 1.0, large amounts of secondary phases were formed, implying a narrow homogeneity range.

Powder x-ray diffraction (XRD) measurements were carried out on a PAN-analytical x-ray diffractometer (Cu-K$\alpha$ radiation) at room temperature. Rietveld refinements were performed with the GSAS package [18]. The composition was checked by a Hitachi S-4800 scanning electron microscope equipped with an Oxford X-Max energy dispersive x-ray (EDX) spectrometer. The EDX measurements gave Mo:Ga:As = 4:19.8(1):0.87(4), which was not far from the nominal composition. The electrical and heat transport measurements were performed on a Quantum Design physical property measurement system (PPMS). Electrical resistivity was measured on a rectangular sample with the standard four-probe technique. Heat capacity was measured with a calorimeter using the relaxation method. Magnetization was measured on a Quantum Design magnetic property measurement system (MPMS). The demagnetization factors were estimated with a similar approach as in the previous study [19].

First-principles calculation was performed with the QUANTUM ESPRESSO package [20], using crystallographic parameters obtained from the XRD refinement. Projector augmented wave pseudopotentials based on the exchange-correlation functionals of PBEsol were adopted [21,22]. Monkhorst–Pack grids of 15×15×5 and 19×19×7 were used to calculate the charge density and the density of states (DOS), respectively. Only the scalar relativistic case was considered.

## 3. Results and discussion

## 3.1. Crystal structure

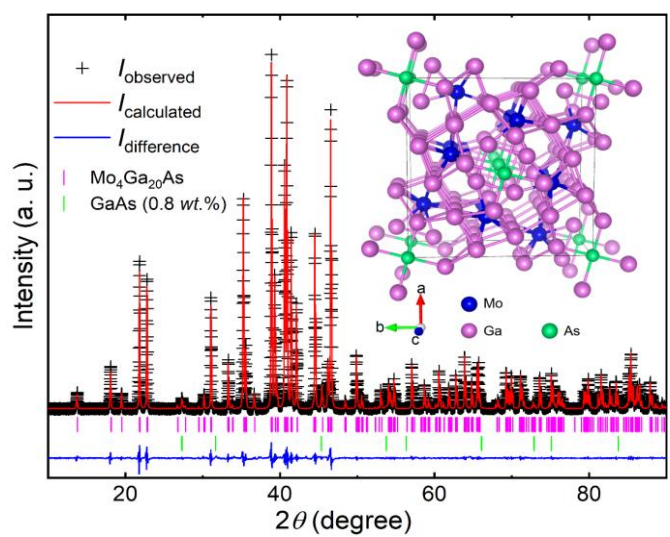

**Figure 1**. Powder x-ray diffraction pattern of $Mo_4Ga_{20}As$ and its Rietveld refinement. Inset shows the crystal structure of $Mo_4Ga_{20}As$.

Figure 1 demonstrates the powder XRD pattern of $Mo_4Ga_{20}As$ and the Rietveld refinement result. According to the refinement, there is about 0.8 *wt.*% of GaAs as the impurity. The crystal structure of $Mo_4Ga_{20}Sb$ was chosen as the starting model of the refinement [15]. As $Mo_4Ga_{20}E$ may host Ga–E substitution [14,15], occupancies on all of the crystallographic sites were allowed to relax in the refinement. However, all these occupancies converged to 1.0, even when we started from fractional values. This means that, although the residual GaAs and EDX results may suggest Ga or As deficiencies, the composition of the sample should be nearly stoichiometric, consistent with our experimental findings during the synthesis procedure. The narrow homogeneity range in $Mo_4Ga_{20}As$ resembles that in $Mo_4Ga_{20}Sb$ [15], which is not surprising as both As and Sb are pnicogens.

Detailed results of the XRD refinement are listed in Table 1. $Mo_4Ga_{20}As$ crystallizes in the space group of $I4/m$ (No. 87), with cell parameters: $a$ = 12.86352(4) Å, $c$ = 5.30031(2) Å. Compared to the previously known $Mo_4Ga_{20}E$ ($E$ = Sn, Sb, S, Se, Te), $c$ for $Mo_4Ga_{20}As$ is larger than that of $Mo_4Ga_{20}Se$ (5.27319 Å) but smaller than that of $Mo_4Ga_{20}Sb$ (5.30905 Å), well reflecting the atomic radius trend in the periodic table, whereas the value of $a$ does not show a regular change compared with the others' [15].

The unit cell of $Mo_4Ga_{20}As$ is illustrated in the inset of Figure 1. The building blocks of the structure are vertex-sharing $Mo@Ga_{10}$ clusters, which are further linked with each other by inter-cluster Ga–Ga and Ga–As bonds.

## 3.2. Superconductivity

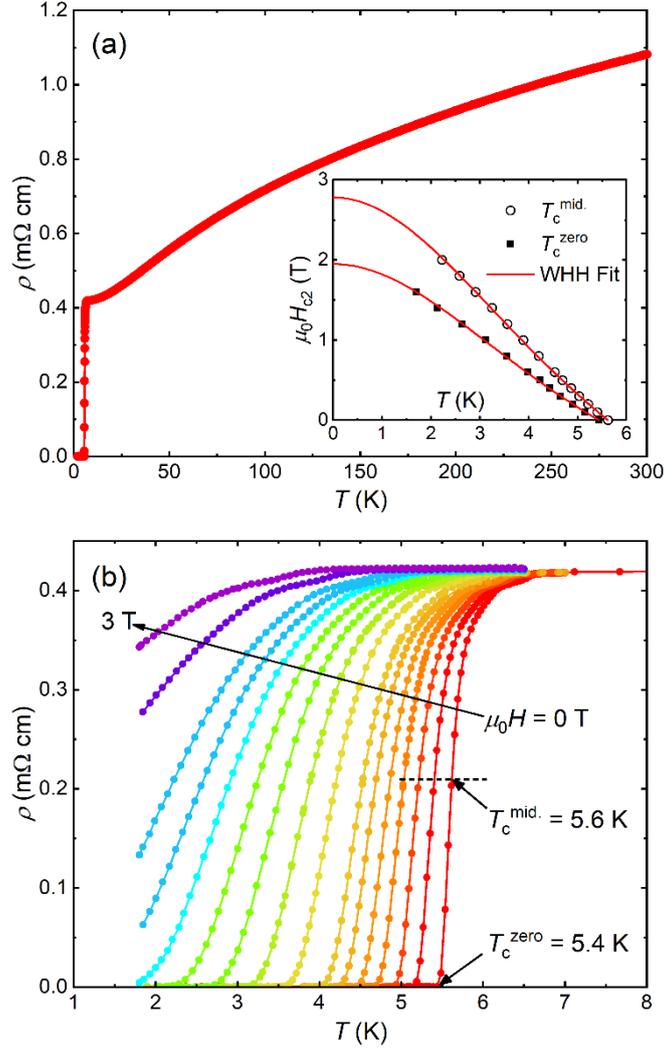

**Figure 2.** (a) Temperature dependence of resistivity of $Mo_4Ga_{20}As$ under zero magnetic field. Inset shows the evolution of $T_c$ upon magnetic fields, solid lines are fits with the WHH model. (b) Superconducting transitions under various magnetic fields up to 3 T.

The temperature dependence of electronic resistivity ($\rho$) of $Mo_4Ga_{20}As$ is shown in Figure 2(a). As the temperature decreasing below 300 K, $\rho$ decreases monotonously, indicating a metallic nature. The low-temperature region of $\rho(T)$ is presented as Figure 2(b), in which the superconducting transition is observed. We notice that there is a small kink at ~ 6.7 K on $\rho(T)$ under zero magnetic field, which is presumably attributed to a tiny amount of $\kappa$-Ga ($T_c$ = 6.7 K) in the sample [23]. Consequently, it is hard to determine the onset value of $T_c$ because of the kink. Nevertheless, we are able to determine the midpoint of the transition ($T_c^{mid.}$), where $\rho(T)$ reaches 50% of that of the normal state, and $T_c^{zero}$, where $\rho(T)$ drops to zero. $T_c^{mid.}$ and $T_c^{zero}$ are thus determined to be 5.6 K and 5.4 K, respectively.

The superconducting transition was gradually suppressed when external magnetic fields were applied, as demonstrated in Figure 2(b). The evolution of $T_c$ upon magnetic field is shown in the inset of Figure 2(a). The upper critical field at zero temperature ($\mu_0H_{c2}(0)$) is determined to be 2.78 T by a Werthamer–Helfand–Hohenberg (WHH) fit of $\mu_0H_{c2}(T)$ from the $T_c^{mid.}$ curve [24].

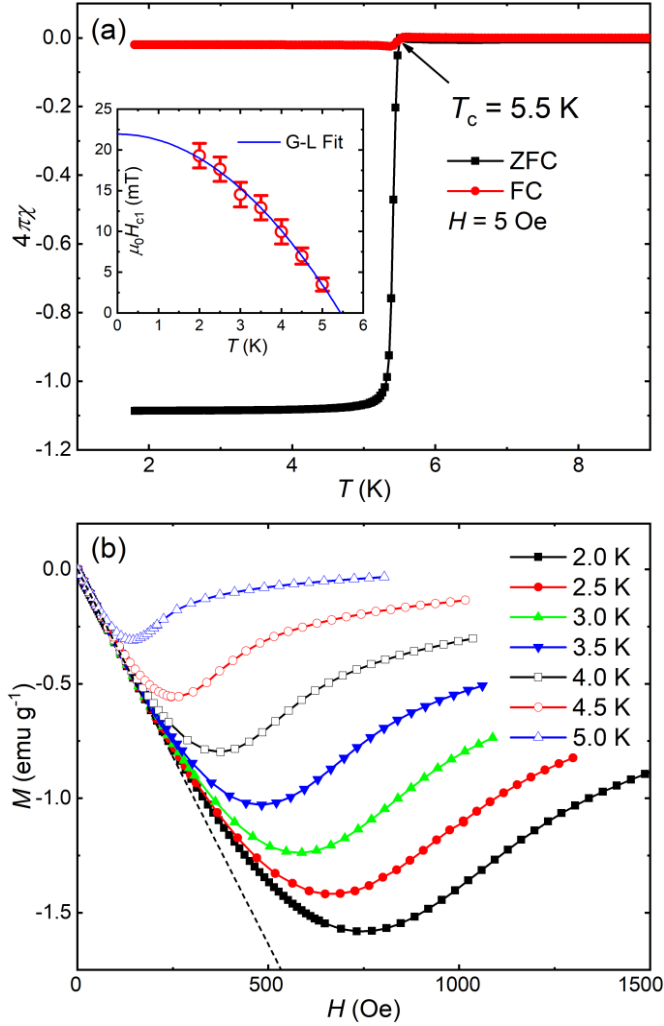

**Figure 3**. (a) Temperature dependence of DC magnetic susceptibility of $Mo_4Ga_{20}As$ under 5 Oe. Inset shows the temperature dependence of lower critical field. (b) Magnetization curves of $Mo_4Ga_{20}As$ under various temperatures below $T_c$. The dash line guides the initial Meissner states.

Superconductivity in $Mo_4Ga_{20}As$ was also confirmed by the magnetization measurements. The temperature dependence of DC magnetic susceptibility ($4\pi\chi$) between 1.8 and 9 K is shown in Figure 3(a). Below 5.5 K, a large diamagnetic signal is observed in the zero-field-cooling (ZFC) run, validating bulk superconductivity. The shielding fraction is slightly larger than 100% because of the error of the sample shape. The superconducting transition width in $4\pi\chi(T)$ is only 0.2–0.3 K, suggesting superior homogeneity of the sample. The diamagnetic signal of the field-cooling (FC) process is much smaller than that of the ZFC process, revealing a substantial pinning effect.

Figure 3(b) shows the isothermal magnetization curves ($M(H)$) of $Mo_4Ga_{20}As$. The initial Meissner states are guided by the dash line. The field where $M(H)$ deviates from the Meissner sate is defined as the lower critical field ($\mu_0H_{c1}$). $\mu_0H_{c1}$ at each temperature is extracted and plotted in the inset of Figure 3(a). $\mu_0H_{c1}(T)$ can be well fitted with the Ginzburg–Landau (G–L) formula [25]: $\mu_0H_{c1}(T) = \mu_0H_{c1}(0)[1 - (T/T_c)^2]$, which yields $\mu_0H_{c1}(0) = 22.0$ mT.

A series of superconducting parameters, including the G–L penetration depth ($\lambda_{GL}$), G–L coherence length ($\xi_{GL}$), G–L parameter ($\kappa_{GL}$), and thermodynamic critical field ($\mu_0H_c(0)$) can be determined from the electrical and magnetic measurements. Methods for the calculation of these parameters can be found in our previous study [26]. The results are summarized in Table 2, from which we note $\kappa_{GL}$ is much larger than $1/\sqrt{2}$, suggesting type-II superconductivity.

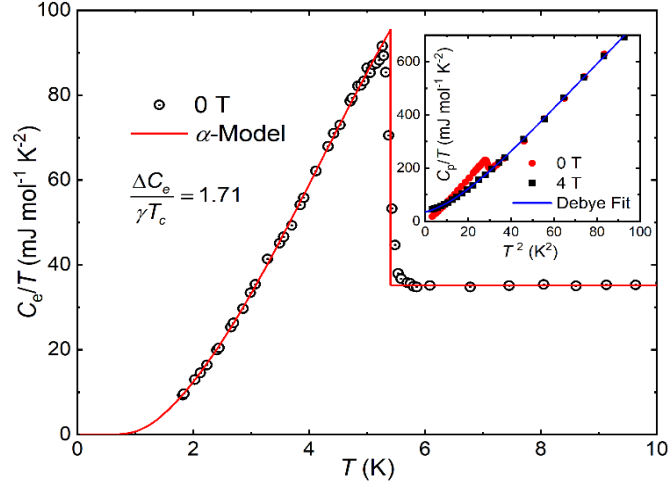

**Figure 4**. Electronic specific heat of $Mo_4Ga_{20}As$ in temperature range of 1.8–10 K. Solid lines indicate the fit with $\alpha$-model. Inset shows the temperature dependence of specific heat ($C_p$) under zero and 4 T magnetic fields. The fit of $C_p$ with Debye model is also shown in the inset.

The temperature dependence of specific heat ($C_p$) from 1.8 to 10 K is shown in the inset of Figure 4. A clear anomaly is observed below 5.4 K, indicating the superconducting transition, which is completely suppressed under a magnetic field of 4 T. $T_c$ determined from the $C_p$ measurements agrees well with those from $\rho(T)$ and $4\pi\chi(T)$. The 4 T curve of $C_p(T)$ can be fitted with the Debye model: $C_p(T) = \gamma T + \beta T^3 + \delta T^5 + \varepsilon T^7$. We notice that the $T^5$ and $T^7$ terms need to be included to describe $C_p(T)$. This might be due to complexity of the phonon spectrum. $\gamma$ and $\beta$ are fitted to be 35.2 mJ mol$^{-1}$ K$^{-2}$ and 2.2 mJ mol$^{-1}$ K$^{-4}$, respectively. Accordingly, the Debye temperature is calculated to be 280 K. This value is almost identical with those of the isostructural compounds $Mo_4Ga_{20}E$ ($E$ = Sb, S, Se, Te, with $\Theta_D$ ranging from 278 to 294 K) [15], which is as expected since they share the same structure and similar compositions. The electron-phonon coupling constant ($\lambda_{ep}$) is thus estimated to be 0.67 with [27]:

$$\lambda_{ep} = \frac{1.04 + \mu^* \ln(\Theta_D / 1.45 T_c)}{(1 - 0.62\mu^*)\ln(\Theta_D / 1.45 T_c) - 1.04}, \quad (1)$$

in which the Coulomb screening parameter $\mu^*$ is set to 0.13. We further estimate the DOS value on the Fermi level ($E_F$): $N(E_F)^{exp.} = 3\gamma/[\pi^2 k_B^2 (1+\lambda_{ep})] = 8.94$ states eV$^{-1}$ per formula unit (f.u.), where $k_B$ is the Boltzmann constant.

The electronic contribution to specific heat ($C_e$) is extracted by: $C_e = C_p|_{0T} - C_p|_{4T} + \gamma$, and is shown in Figure 4. The normalized $C_e$ change at $T_c$ ($\Delta C_e/\gamma T_c$) is 1.71, which is larger than that of the Bardeen–Cooper–Schrieffer (BCS) weak coupling ratio (1.43). $C_e$ at the superconducting state can be well fitted with the $\alpha$-model [28], yielding a zero-temperature gap value $\Delta_0$ of 0.86 meV. Thus $\Delta_0/k_B T_c = 1.85$, evidencing strong electron-phonon coupling. The thermodynamic and superconducting parameters of $Mo_4Ga_{20}As$ are listed in Table 2.

*3.3. First-principles calculation*

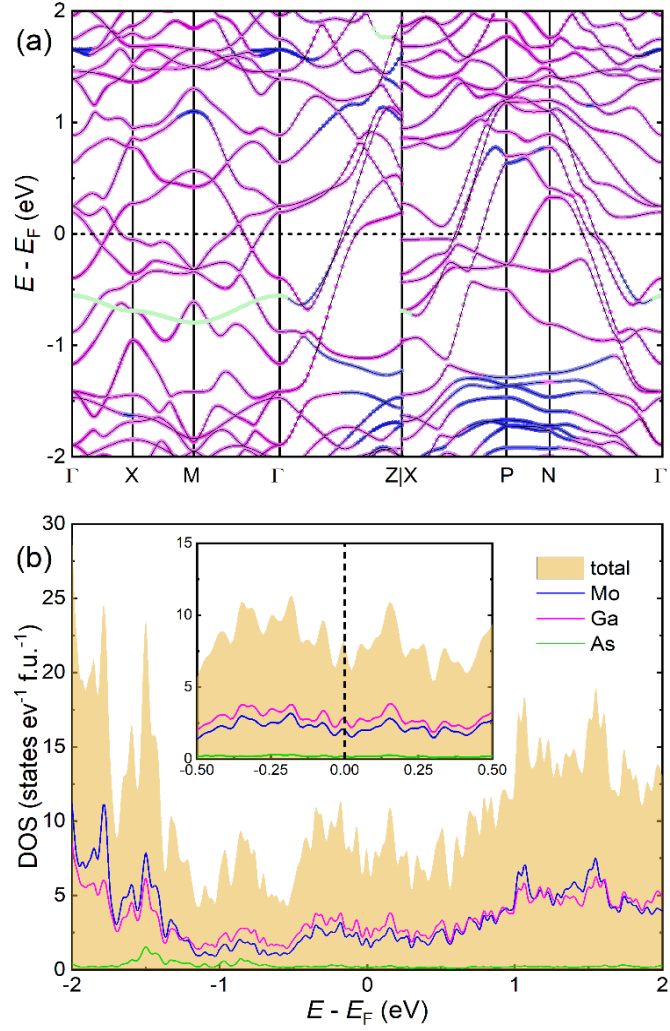

**Figure 5**. (a) Calculated electronic band structure. The blue, magenta, and green dots represent the contributions from Mo, Ga, and As atoms, respectively. (b) Calculated density of states. Inset shows a zoom-in near the Fermi level.

Results from the density-functional theory (DFT) calculations are shown in Figure 5. There are multiple bands crossing $E_F$, which is consistent with the metallic behaviour. The energy dispersions between –1 eV and 1 eV are generally large, while flat bands appear below ~ –1.2 eV and above ~ 1.5 eV. Such a feature is similar with the cases in isostructural Mo$_4$Ga$_{20}$E (E = Sn, Sb, S, Se, Te) compounds [14,15]. Likewise, the states near $E_F$ are also dominated by the Mo-4$d$ and Ga-4$p$ orbitals with minor contributions from the As-4$p$ orbitals, as shown in the band structure and DOS plots.

From the inset of Figure 5(b), one may notice that $E_F$ locates at a small peak of DOS. In addition, we obtain a theoretical DOS at $E_F$: $N(E_F)^{calc.}$ = 8.16 states eV$^{-1}$ f.u.$^{-1}$, which is close to the experimental value. We also note that $N(E_F)$ could be enhanced with carrier doping, since the neighboring DOS peaks are higher than that on $E_F$. This means that slight electron or hole doping into Mo$_4$Ga$_{20}$As may lead to an increase of $T_c$ (assuming a rigid band).

*3.4. Discussion*

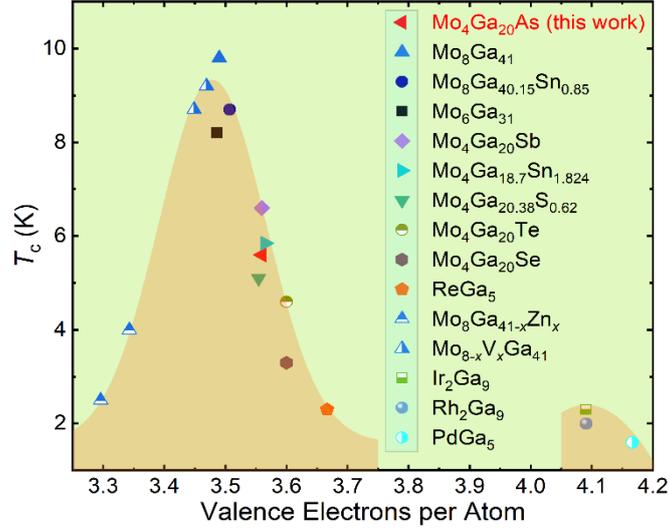

**Figure 6**. Evolution of $T_c$ upon valence electrons per atom ($e/a$) in superconductors with endohedral gallium clusters.

In Figure 6, the $T_c$ values of transition metal (*TM*) gallide superconductors, including the Mo$_4$Ga$_{20}E$ family [14,15], the Mo$_8$Ga$_{41}$ family [12,29], Mo$_6$Ga$_{31}$ [13], Rh$_2$Ga$_9$/Ir$_2$Ga$_9$ [10], ReGa$_5$ [30], and PdGa$_5$ [19], are summarized. Although crystallizing in different structures, they share much in common. For example, the compositions are close to *TM*Ga$_5$, and most of the crystal structures consist of *TM*@Ga$_{10}$ endohedral clusters (except for Rh$_2$Ga$_9$ and Ir$_2$Ga$_9$, which are built with *TM*@Ga$_9$ clusters).

Interestingly, $T_c$s in these gallides form dome-like dependence upon the valence electrons per atom ($e/a$), with one pronounced peak at $e/a \sim 3.5$ (for Mo$_8$Ga$_{41}$, $T_c = 9.8$ K) and another small dome at $e/a \sim 4.1$. Such evolution of $T_c$ also confirms previous findings [15,19,30]. A dome-like dependence, known as the Matthias' rule [31], was also observed in the W$_5$Si$_3$-type superconductors [32]. The regular evolution of $T_c$ upon $e/a$ is closely related to the change of $N(E_F)$ due to different valence electron numbers. Indeed, the theoretical $N(E_F)$ is reduced from 0.456 (for Mo$_8$Ga$_{41}$ [12]) to 0.326 (for Mo$_4$Ga$_{20}$As), and further to 0.217 (for ReGa$_5$ [30]) states eV$^{-1}$ per atom with $e/a$ increasing, consistent with the trend of $T_c$. Simultaneously, the electron-phonon coupling decreases, resulting in a transition from a strong-coupling regime to a moderate or weak-coupling one. For a BCS superconductor, $\lambda_{ep}$ is proportional to $N(E_F)$ [27]. These results thus imply that $N(E_F)$ is a significant determinant of electron-phonon coupling in endohedral cluster superconductors, regardless of structural discrepancies. We should note that the qualitative $T_c$–$N(E_F)$ relation holds true only for compounds with relatively small $e/a$. This is the reason we plot Rh$_2$Ga$_9$, Ir$_2$Ga$_9$, and PdGa$_5$ in a separated dome in Figure 6. Nevertheless, tuning $e/a$ via doping should be an effective method to change $T_c$ in the endohedral cluster superconductors.

## 4. Conclusions

Mo$_4$Ga$_{20}$As, a new member of the Mo$_4$Ga$_{20}E$ family, was successfully synthesized. The crystal structure consists of endohedral Mo@Ga$_{10}$ clusters. Mo$_4$Ga$_{20}$As is a type-II superconductor with $T_c$ = 5.6 K and possible strong electron-phonon coupling. The states at the Fermi level are mainly contributed by the Mo-4$d$ and Ga-4$p$ orbitals, as revealed by the first-principles calculations. Our work confirms the chemical flexibility of the Mo$_4$Ga$_{20}$$E$ family, hinting that more novel compounds with different $E$ elements could also be obtained. In particular, carrier doping may lead to new superconductors with higher $T_c$ values in this structural family.

**Acknowledgements**
This work was supported by the National Natural Science Foundation of China (Grant Nos. 12074414 and 11774402), the National Key Research and Development Program of China (Grant Nos. 2018YFA0704200 and 2021YFA1401800), and the Strategic Priority Research Program of Chinese Academy of Sciences (Grant No. XDB25000000).

**Table 1.** Crystallographic parameters of $Mo_4Ga_{20}As$ obtained from Rietveld refinement of XRD. $R_p$ = 3.22%, $R_{wp}$ = 2.18%, $Z$ = 2, space group $I4/m$ (No. 87), $a$ = 12.86352(4) Å, $c$ = 5.30031(2) Å, $V$ = 877.04(2) Å$^3$. $U_{eq}$ is one-third of the trace of the orthogonalized $U_{ij}$ tensor.

| Atom | Wyckoff Position | $x$ | $y$ | $z$ | $U_{eq}$ (0.01 Å$^2$) | Occupancy |
|---|---|---|---|---|---|---|
| Mo | 8$h$ | 0.4188(1) | 0.23172(8) | 0 | 0.79(3) | 1.0 |
| Ga1 | 8$h$ | 0.3654(1) | 0.4293(2) | 0 | 3.12(15) | 1.0 |
| Ga2 | 8$h$ | 0.0530(2) | 0.1998(1) | 0 | 2.03(13) | 1.0 |
| Ga3 | 16$i$ | 0.3631(1) | 0.0656(1) | 0.2463(4) | 2.04(10) | 1.0 |
| Ga4 | 8$f$ | 0.25 | 0.25 | 0.25 | 2.03(11) | 1.0 |
| As | 2$a$ | 0 | 0 | 0 | 2.68(17) | 1.0 |

**Table 2.** Normal state and superconducting parameters of $Mo_4Ga_{20}As$.

| Parameter (unit) | Value |
|---|---|
| $T_c^{mid.}$ (K) | 5.6 |
| $T_c^{zero}$ (K) | 5.4 |
| $\mu_0H_{c1}(0)$ (mT) | 22.0 |
| $\mu_0H_{c2}(0)$ (T) | 2.78 |
| $\mu_0H_c(0)$ (T) | 0.15 |
| $\lambda_{GL}$ (nm) | 153.3 |
| $\xi_{GL}$ (nm) | 10.9 |
| $\kappa_{GL}$ | 14.1 |
| $\gamma$ (mJ mol$^{-1}$ K$^{-2}$) | 35.2 |
| $\beta$ (mJ mol$^{-1}$ K$^{-4}$) | 2.2 |
| $\Theta_D$ (K) | 280 |
| $\Delta C_e/\gamma T_c$ | 1.71 |
| $\Delta_0$ (meV) | 0.86 |
| $\lambda_{ep}$ | 0.67 |
| $N(E_F)^{exp.}$ (states eV$^{-1}$ f.u.$^{-1}$) | 8.94 |
| $N(E_F)^{calc.}$ (states eV$^{-1}$ f.u.$^{-1}$) | 8.16 |